# Magnetic proximity in a van der Waals heterostructure of magnetic insulator and graphene

Bogdan Karpiak[1], Aron W. Cummings[2], Klaus Zollner[3], Marc Vila[2,4], Dmitrii Khokhriakov[1], Anamul Md Hoque[1], André Dankert[1], Peter Svedlindh[5], Jaroslav Fabian[3], Stephan Roche[2,6], Saroj P. Dash[1,7]

[1] *Department of Microtechnology and Nanoscience, Chalmers University of Technology, SE-41296, Göteborg, Sweden*
[2] *Catalan Institute of Nanoscience and Nanotechnology (ICN2), CSIC and The Barcelona Institute of Science and Technology, Campus UAB, Bellaterra, 08193 Barcelona, Spain*
[3] *Institute for Theoretical Physics, University of Regensburg, 93040 Regensburg, Germany*
[4] *Department of Physics, Universitat Autònoma de Barcelona, Campus UAB, Bellaterra, 08193 Barcelona, Spain*
[5] *Division of Solid-State Physics, Department of Engineering Science, Uppsala University, Box 534, SE-75121 Uppsala, Sweden*
[6] *ICREA—Institució Catalana de Recerca i Estudis Avançats, 08010 Barcelona, Spain*
[7] *Graphene center, Chalmers University of Technology, SE-41296, Göteborg, Sweden*

**Abstract**

Engineering two-dimensional material heterostructures by combining the best of different materials in one ultimate unit can offer a plethora of opportunities in condensed matter physics. Here, in the van der Waals heterostructures of the ferromagnetic insulator $Cr_2Ge_2Te_6$ and graphene, our observations indicate an out-of-plane proximity-induced ferromagnetic exchange interaction in graphene. The perpendicular magnetic anisotropy of $Cr_2Ge_2Te_6$ results in significant modification of the spin transport and precession in graphene, which can be ascribed to the proximity-induced exchange interaction. Furthermore, the observation of a larger lifetime for perpendicular spins in comparison to the in-plane counterpart suggests the creation of a proximity-induced anisotropic spin texture in graphene. Our experimental results and density functional theory calculations open up opportunities for the realization of proximity-induced magnetic interactions and spin filters in 2D material heterostructures and can form the basic building blocks for future spintronic and topological quantum devices.

**Introduction**

Topological quantum states of matter and spintronics have considerable interest in the field of condensed matter physics for applications in low-power electronics without the application of an external magnetic field[1,2]. The generation of magnetic exchange interaction and strong spin-orbit coupling in two-dimensional (2D) Dirac materials such as graphene is expected to result in the emergence of quantum anomalous Hall state and topologically protected chiral spin textures[3,4]. Graphene, having excellent charge and spin transport properties, is a suitable atomically-thin 2D material to create proximity-induced effects when placed in heterostructures with other functional materials[5–7]. Proximity-induced magnetic effects have been investigated in 2D semiconductors on graphene in heterostructures with ferromagnetic semiconductors[8] and magnetic oxides[9–16]. However, there are severe challenges in producing a sizable out-of-plane exchange interaction in magnetic oxide-graphene structures due to the in-plane magnetic anisotropy of the oxide-based magnetic insulators used so far with graphene[13].

To create significant proximity-induced magnetic interaction in graphene, the use of a magnetic insulator with perpendicular magnetic anisotropy is desired. Additionally, for a good interface, atomically-flat layered 2D-material-based van der Waals heterostructures (vdWh) are ideal[1–4]. Recently, 2D magnetic materials were successfully prepared via exfoliation down to the monolayer limit, showing intrinsic long-range Heisenberg ferromagnetic order[17–19] that is tunable by application of a gate voltage[20–22]. Additionally, giant tunneling magnetoresistance, spin filtering and magnon-assisted tunneling phenomenon have been demonstrated using such 2D magnetic materials[23–27]. Recently, heterostructures of such 2D magnetic materials with 2D semiconductors[8], topological insulators[28], semimetals[29] and graphene[27] have also been investigated. Theoretical predictions indicate that heterostructures of such 2D magnetic insulators with graphene are expected to produce a large exchange splitting in the graphene layer, enabling the emergence of a topological quantum phase[30–32].

Here, we use van der Waals heterostructures of graphene with the ferromagnetic insulator $Cr_2Ge_2Te_6$ (CGT) having a perpendicular magnetic anisotropy to demonstrate a proximity-induced magnetic exchange interaction, which is revealed by the temperature-dependent splitting of Hanle spin precession signals. The observed anisotropic spin relaxation indicates a possible nontrivial spin texture imprinted in graphene by proximity to CGT. The demonstration of out-of-plane proximity-induced magnetism induced in graphene is a crucial step towards realizing more exotic electronic states in 2D material heterostructures.

## Results

The van der Waals heterostructures are prepared by dry-transferring CGT flakes (~30 nm in thickness) onto chemical vapor deposited (CVD) graphene on a SiO$_2$/n-Si substrate in a cleanroom environment within one minute after exfoliation. The atomic structure of the CGT flakes consists of van der Waals layers of Cr atoms sandwiched by Te and Ge atoms with an interlayer distance of ~6.9 Å (Fig. 1a)[33,34]. The choice of CGT is motivated by its layered structure, insulating behavior[30,35], and perpendicular magnetic anisotropy[17]. CGT is expected to induce a magnetic exchange interaction in graphene, splitting the spin-degenerate graphene bands by a characteristic exchange $\Delta E_{\text{ex}}$ (Fig. 1b). The Raman spectrum of the CGT flake in a heterostructure with graphene is shown in Fig. 1c. It contains the phonon mode at 108 cm$_{-1}$, which is characteristic of a CGT flake with more than one layer[36]. We do not observe any broad peak at 121 cm$_{-1}$, characteristic of an oxidized CGT surface (Supplement Fig. S1) after long exposure to the ambient atmosphere, indicating absence of substantial oxidation in freshly cleaved CGT flakes. The Raman spectrum of the CVD deposited graphene channel (Fig. 1c) contains a 2D peak (2642 cm$_{-1}$) of stronger intensity than the G peak (1600 cm$_{-1}$), indicating monolayer thickness of graphene[37]. Magnetic characterization of the bulk CGT crystal by SQUID magnetometry shows perpendicular magnetic anisotropy of the material (Fig. 1d). From the temperature dependence of the magnetization of the bulk CGT crystal under a perpendicularly applied magnetic field of 100 mT (Fig. 1e), one can observe two magnetic ordering temperatures. By fitting the inverse magnetic susceptibility according to the Curie-Weiss law $\frac{1}{\chi} = \frac{(T-Tc)^{\gamma}}{C}$ (inset in Fig. 1e), values of the transition temperatures of ~65 K ($\gamma = 1$) and ~204 K ($\gamma \approx 1.6$) were obtained. The former shows a more distinct magnetization enhancement with deceasing temperature. The magnetic moment $m$ of the high temperature ($T_c$~204 K) phase is however two orders of magnitude smaller than that of the main magnetic phase ($T_c$~65 K). The presence of a magnetic transition at higher temperature (Fig. 1e and Supplement Fig. S2) could be due to imperfections in the lattice structure or an extra magnetic phase leading to a magnetic transition at higher temperature in the regions of the material with broken stoichiometry and lattice defects.

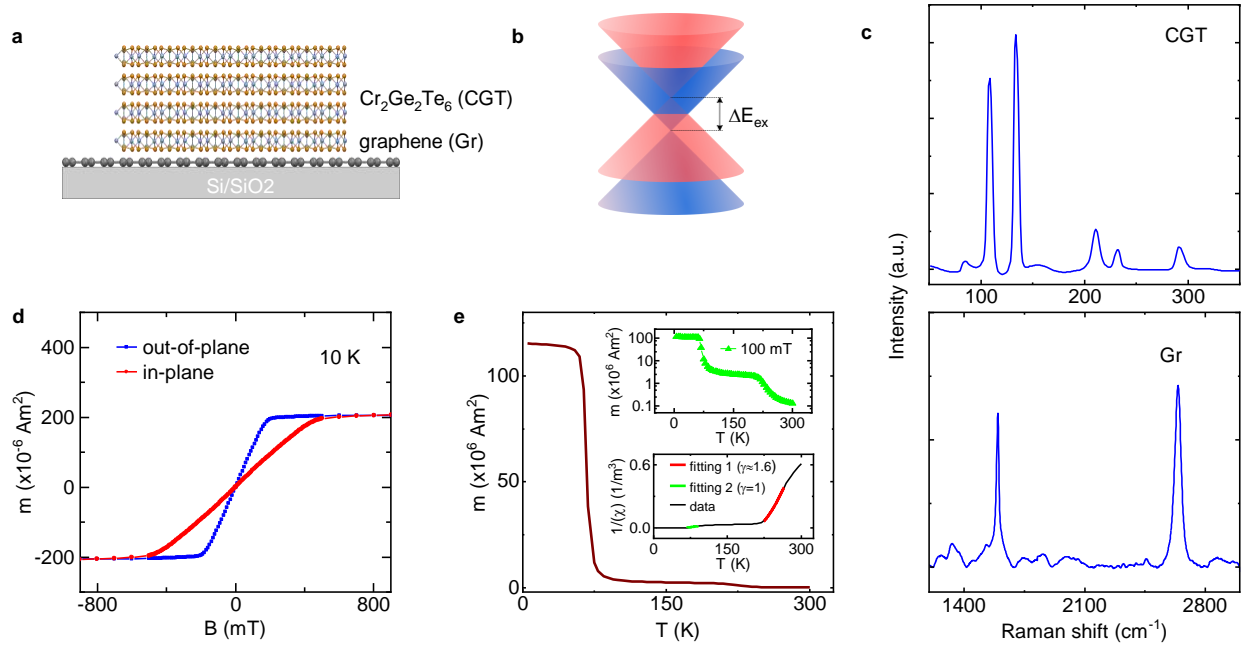

*Figure 1. Characterization of Cr$_2$Ge$_2$Te$_6$ and its heterostructure with graphene. **a**,* Schematic of atomic structure of CGT/graphene heterostructure. ***b**,* Schematic of graphene band structure in proximity with CGT in the vicinity of the Dirac point. Graphene can acquire magnetic order with a splitting of spin-degenerate bands by an exchange $\Delta E_{\mathrm{ex}}$. ***c**,* Raman spectrum of CGT (top panel) and graphene (bottom panel) in the heterostructure. ***d**,* Field dependence of magnetic moment (m) of bulk CGT crystal as a function of magnetic field at 10 K for both in-plane and out-of-plane magnetic field directions. ***e**,* Temperature dependence of magnetic moment m of CGT under perpendicular magnetic field of 100 mT. Top inset shows the same data as in main panel but with log scale on vertical axis. Bottom inset shows inverse magnetic susceptibility as a function of temperature fitted with the Curie-Weiss law.

## Magnetic proximity effect in graphene-Cr$_2$Ge$_2$Te$_6$ heterostructures

The device schematic and an optical microscope image of a nanofabricated device consisting of a graphene–CGT heterostructure channel are shown in Fig. 2a and 2b (see Methods for details about the device fabrication process). The ferromagnetic tunnel contacts of Co/TiO$_2$ on graphene are used for injection and detection of the spin-polarized current in the heterostructure channel, with three-terminal contact resistances of 3-6 kΩ at room temperature (Supplement Fig. S3). In such graphene-CGT heterostructure devices, we probe the proximity-induced exchange interaction by employing spin transport and Hanle precession measurements. The insulating behavior of CGT (two-terminal resistance ∼60 MΩ) allows charge carriers to flow mostly in the graphene layer with a field-effect mobility of ∼2400 cm$^2$V$^{-1}$s$^{-1}$ in the heterostructure channel (Supplement Fig. S4).

First, spin-valve and Hanle precession measurements were carried out at room temperature (300 K), well above the $T_c$ of CGT, to check the functionality of the heterostructure devices in a nonlocal (NL) configuration (see Fig. 2a). A NL spin resistance change $\Delta R_{NL} = \Delta V_{NL}/I \approx 5.5$ mΩ was measured at 300 K while sweeping the in-plane magnetic field ($B_\parallel$) along the easy axis of the Co contacts in a device with channel length $L = 6.9$ μm (Fig. 2c). Next, Hanle measurements were conducted in the NL configuration but with an out-of-plane applied magnetic field ($B_\perp$), which causes in-plane Larmor precession of spins as they propagate through the heterostructure channel. Such measurements at room temperature (Fig. 2d) for both up and down magnetic field sweeps showed conventional Hanle signals without any special features[38]. By fitting the data with the solution of the Bloch equation that describes spin dynamics in the channel with an applied $B_\perp$[38], the spin lifetime $\tau_s = 244 \pm 32$ ps, diffusion coefficient $D_s = 0.019 \pm 0.005$ m2s-1 and diffusion length $\lambda = 2.1 \pm 0.4$ μm were obtained.

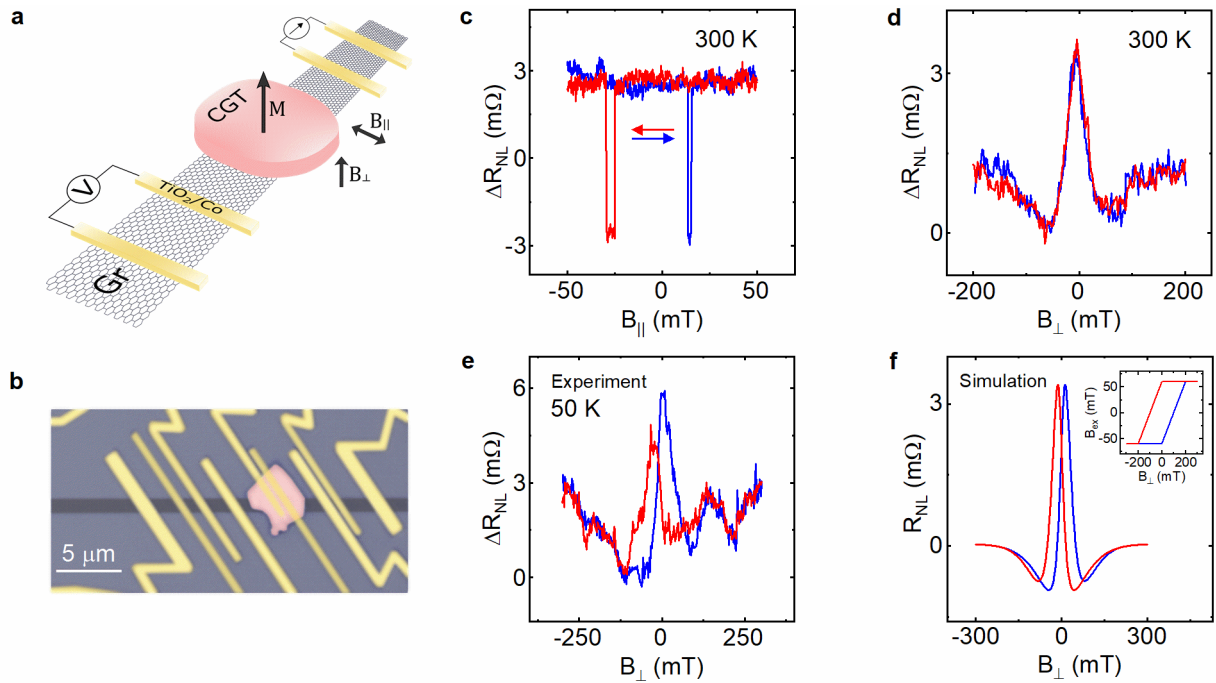

*Figure 2. Magnetic proximity effect in graphene-Cr2Ge2Te6 heterostructures. a,b,* Schematic and false-color optical microscope picture of graphene-CGT heterostructure with ferromagnetic tunnel contacts of TiO2 (1 nm)/Co (60 nm). The scale bar is 5 μm. *c,d,* Spin valve and Hanle spin precession signals measured in graphene-CGT heterostructure for in-plane and out-of-plane magnetic field directions, respectively, at 300 K, above the Curie temperature of CGT. The up and down magnetic field sweep directions are represented by blue and red colors. *e,* The measured Hanle signal at 50K, below the Curie temperature of CGT. *f,* Simulated Hanle signal for typical channel parameters (see Supplementary note 1 for details) with partial overlap of the graphene by CGT. The inset shows the assumed hysteresis for CGT flake in the simulations. The overlap area of the graphene-CGT heterostructure in the device channel is ~60 %.

At low temperatures (50 K), below the Curie point of CGT (Fig. 2e), one can notice two distinctive features of the measured Hanle signal: a shift of the two Hanle peaks with respect to each other and asymmetry of the Hanle peak. Such observations are indicative of an out-of-plane magnetism induced in the graphene by proximity to CGT. These features are also seen in simulations (Fig. 2f and Supplementary note 1) obtained from the solution of the Bloch equation that considers a magnetic exchange interaction in graphene due to the influence of the CGT flake[39]. The CGT flake can induce ferromagnetism in the underlying graphene via the exchange field $B_{ex}$, which adds to the external magnetic field, $B_\perp \rightarrow B_\perp + B_{ex} + B_s$, and thus modifies the Hanle spin precession signal in the heterostructure channel. While the stray fields $B_s$ do contribute to the spin precession in the channel in a similar way as $B_{ex}$, the observed Hanle signals are mainly shaped by the contributions from externally applied and exchange fields (see Supplementary note 2). The peak splitting between the two Hanle sweeps can then originate from hysteretic behavior[20] of the proximity-induced magnetic exchange fields in graphene. This alters the position of the Hanle peak corresponding to zero total field, which, due to hysteresis, is obtained at different values of the applied magnetic field depending on the sweep direction.

Meanwhile, the asymmetry of the Hanle peak arises from the spatial inhomogeneity of the proximity-induced exchange field; a finite graphene-CGT interfacial region leads to an asymmetry, while CGT covering the entire graphene flake would result in a symmetric Hanle curve (see Supplementary note 1). Since hysteresis gives opposite shifts of the exchange field depending on the sweep direction, this is reflected in the opposite asymmetry of the measured Hanle signals (Fig. 2e,f). The measurements in the two devices with different CGT/graphene channel overlaps by the CGT flake confirm decreased asymmetry with increased channel overlap (Supplementary note 3), which is in agreement with simulation results (Supplementary note 1).

To further elucidate the proximity-induced magnetic interaction in graphene-CGT heterostructures, the Hanle spin precession measurements were performed at several different temperatures in a systematic manner (Fig. 3a). The measured data reveal a rapid decrease of the nonlocal spin signal amplitude ($R_{NL}$) with increasing temperature in comparison to the temperature dependence of pristine graphene (Fig. 3b)[38], followed by a saturation of the signal above the curie temperature of the CGT. The rapid spin signal amplitude decay with temperature in the range of $T<T_c$, compared to pristine graphene[38], can be attributed to fluctuating proximity-induced magnetic exchange fields due to random fluctuations of the magnetization of the CGT flake, which become more pronounced approaching the magnetic ordering temperature of CGT[15]. Such fluctuations cause random changes of the proximity-induced exchange field and, hence, the effective field that acts on the propagating spins, leading to enhanced spin relaxation[40,41]. Furthermore, the separation between the peaks for opposite Hanle sweeps vanishes at elevated temperatures (Fig. 3c).

The presence of Hanle peak shifts indicates that the proximity-induced exchange field in graphene persists above the Curie temperature for bulk CGT[28] (~65 K) up to at least 100 K. Additionally, one can deconvolute the measured Hanle signal into symmetric and asymmetric components, where the latter contains information on additional rotation of spins in the channel[42] with respect to normal precession caused by the applied perpendicular magnetic field. Such rotation can originate e.g. from the modified spin texture of graphene under the CGT flake caused by the proximity to the latter. From Fig. 3d,e one can see that the asymmetric contribution to the signal also vanishes with increasing temperature in a comparable way as in Fig. 3c. This suggests the same origin for peak splitting and asymmetry - both can arise from the proximity-induced exchange field in the graphene channel under the flake, which persists at $T > T_c$ of bulk CGT and decays with increasing temperature, while being completely absent at temperatures above 165 K. Fitting the symmetric component of the Hanle signal with the solution of the Bloch equation for spin diffusion in a homogeneous channel gives values of the in-plane spin lifetimes $\tau_\parallel$ in the range of 90-280 ps; the comparably large spread of values results due to a large number of free fitting parameters (see Supplementary note 4). However, most fits resulting in reasonable spin transport parameters yielded an exchange field on the order of a few 10's of mT.

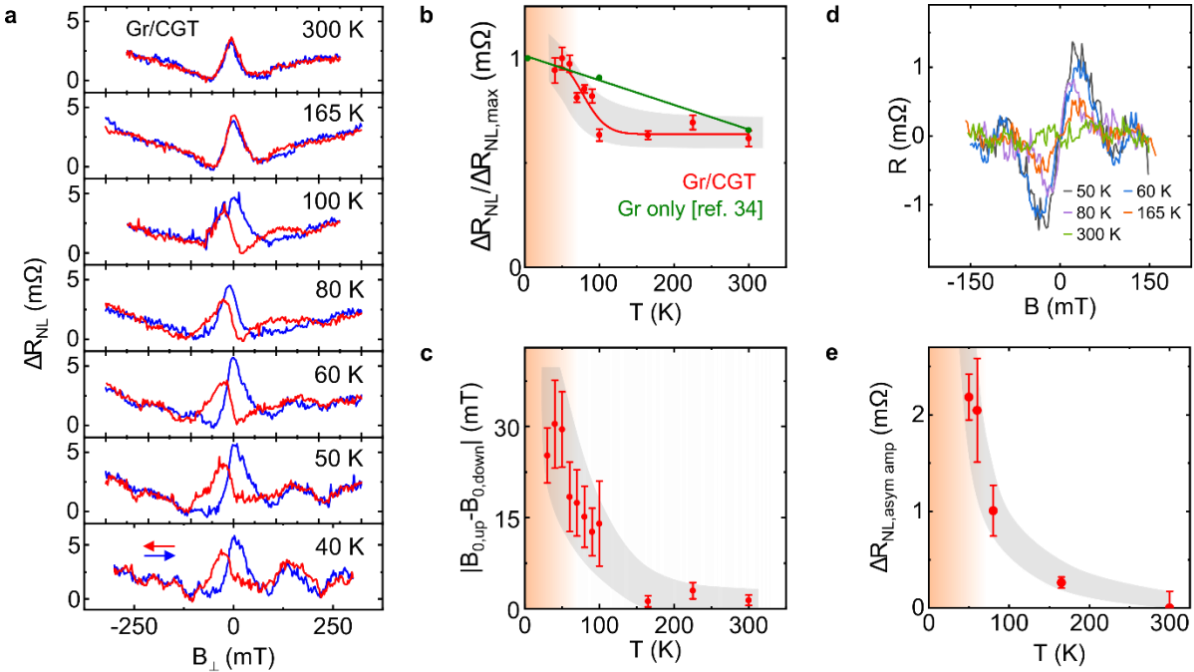

*Figure 3. Temperature dependence of spin signal in graphene-Cr$_2$Ge$_2$Te$_6$ heterostructures. **a,** Nonlocal Hanle spin precession measurements in the graphene-CGT heterostructure as a function of perpendicular magnetic field for up (blue) and down (red) sweeps at different temperatures. **b,** Normalized nonlocal Hanle signal amplitudes (ΔR$_{NL}$) as a function of temperature for a graphene-CGT device and for a 4-μm CVD*

*graphene channel38 (green). **c,** Magnitude of the horizontal separation between positions of nonlocal Hanle signal peaks for up ($B_{0,\text{up}}$) and down ($B_{0,\text{down}}$) sweeps as a function of temperature. **d,** Asymmetric part of the measured Hanle signal at different temperatures. **e,** Amplitude of the asymmetric signal from **d** as a function of temperature. The solid lines in **b** and grey regions in **b,c,e** are guides to the eye.*

**Anisotropic spin relaxation in graphene-CGT insulator heterostructures**

The proximity-induced exchange interactions in the graphene-based heterostructures can also result in spin relaxation anisotropy. Here we observe spin relaxation anisotropy in graphene-CGT heterostructures by employing the same NL Hanle measurement configuration. However, here the field sweep range is broader, allowing the injection and detection of both in-plane and out-of-plane spins in different field ranges. In-plane spins are injected and detected at low field values when the contacts are magnetized in-plane at magnetic fields $B_\perp$<0.4 T (Fig. 4a). With increasing out-of-plane field the magnetization of the contacts starts to rotate to the out-of-plane direction, reaching saturation at $B_\perp$>2 T (Fig. 4b). When spins relax with different rates in the in-plane or out-of-plane direction, different amplitudes of the signal are observed for the corresponding magnetic field ranges (Fig. 4c). Therefore, the in-plane spin lifetime $\tau_\parallel$ is estimated from low $B_\perp$-field Hanle data, while the out-of-plane spin lifetime $\tau_\perp$ is extracted from large out-of-plane $B_\perp$-field measurements. As shown in Fig. 4d, in the graphene-CGT heterostructure the spin signal magnitudes at $B_\perp$=0 for in-plane spins ($\Delta R_{\text{NL}}^\parallel$) and at $B_\perp$=2.5 T for spins perpendicular to plane ($\Delta R_{\text{NL}}^\perp$) show strong anisotropy, with $\Delta R_{\text{NL}}^\perp > \Delta R_{\text{NL}}^\parallel$ and with a ratio $\Delta R_{\text{NL}}^\perp/\Delta R_{\text{NL}}^\parallel \sim 10$. The spin lifetime anisotropy $r \equiv \tau_\perp/\tau_\parallel$ is obtained from $\Delta R_{\text{NL}}^\perp/\Delta R_{\text{NL}}^\parallel = \sqrt{r}e^{-(L/\lambda_\parallel)(\frac{1}{\sqrt{r}}-1)}$, where $L$ is the channel length and $\lambda_\parallel$ is the spin diffusion length for in-plane spins43. From this expression we find a strong anisotropy, with $\tau_\perp$ being 3.9 times the value of $\tau_\parallel \sim 150$ ps (assuming homogeneous channel characteristics). Such observations are in contrast with the isotropic spin relaxation in pristine graphene channels, which have a value of r close to 1 (0.94±0.01 in Fig. 4e).

      The observed anisotropic spin relaxation in graphene-CGT heterostructures could occur, for example, if spin relaxation is dominated by the so-called valley-Zeeman spin-orbit coupling (SOC), similar to the case of graphene on transition metal dichalcogenides44,45. Valley-Zeeman SOC ($\lambda_{\text{VZ}}$) generates an out-of-plane spin texture that, in the presence of strong inter-valley scattering ($\tau_{\text{iv}}$), quickly relaxes the in-plane spins and results in large anisotropy. To determine the type and magnitude of SOC induced in graphene by CGT, we perform density functional theory (DFT) simulations of the graphene-CGT heterostructure (see Methods and Supplementary note 5). By evaluating the band splitting at both K and K' valleys, we obtain a Rashba SOC $\lambda_{\text{R}}$ = 0.253 meV and $\lambda_{\text{VZ}}$ = 0.113 meV. Assuming $\tau_{\text{iv}} = (5-10)\tau_{\text{p}}$, where $\tau_{\text{p}}$ is the momentum scattering time, we obtain an anisotropy of the order $r = (\lambda_{\text{VZ}}/\lambda_{\text{R}})^2(\tau_{\text{iv}}/\tau_{\text{p}}) + 1/2 = 1.5 - 2.5$.

While SOC can induce a spin lifetime anisotropy $r > 1$, the temperature dependence of the spin signal in Fig. 3b suggests that magnetic exchange fluctuations are playing a role in the spin relaxation. In this scenario, an apparent spin relaxation anisotropy can also emerge. In case of the exchange fluctuation mechanism the spins are dephased by the fluctuations, but if magnetic field is applied the fluctuations are suppressed. This is different for spins parallel and perpendicular to the applied magnetic field, which results in an increase of anisotropy with increasing out-of-plane magnetic field $B_\perp$. Assuming that exchange fluctuations are the sole source of spin relaxation in the graphene-CGT devices, we find that the spin relaxation is nearly completely isotropic ($r \approx 1.002$) at $B_\perp=0$ and becomes anisotropic with increasing values of $B_\perp$ (see Supplementary note 6). This analysis indicates that if the spin relaxation is dominated by exchange fluctuations, the large anisotropy seen in Fig. 4d can be driven by the external field $B_\perp$ and may not be intrinsic to the graphene/CGT interface. This is because the perpendicular proximity-induced exchange field $B_{ex}$, while sufficiently large to see in Hanle measurements, is not large enough to suppress exchange fluctuations in graphene. It is still unclear whether spin relaxation is dominated by SOC, exchange fluctuations, or a combination of the two. However, considering that according to DFT calculations the exchange splitting is expected to be an order of magnitude stronger than SOC (see discussion for Supplement Figure S10), one can expect the origin of the experimentally observed anisotropy of spin relaxation to be mainly caused by the exchange fluctuations. Figure 3b points to at least some contribution of exchange fluctuations, while the anisotropy measurement in Fig. 4d can be explained by either of the mechanisms.

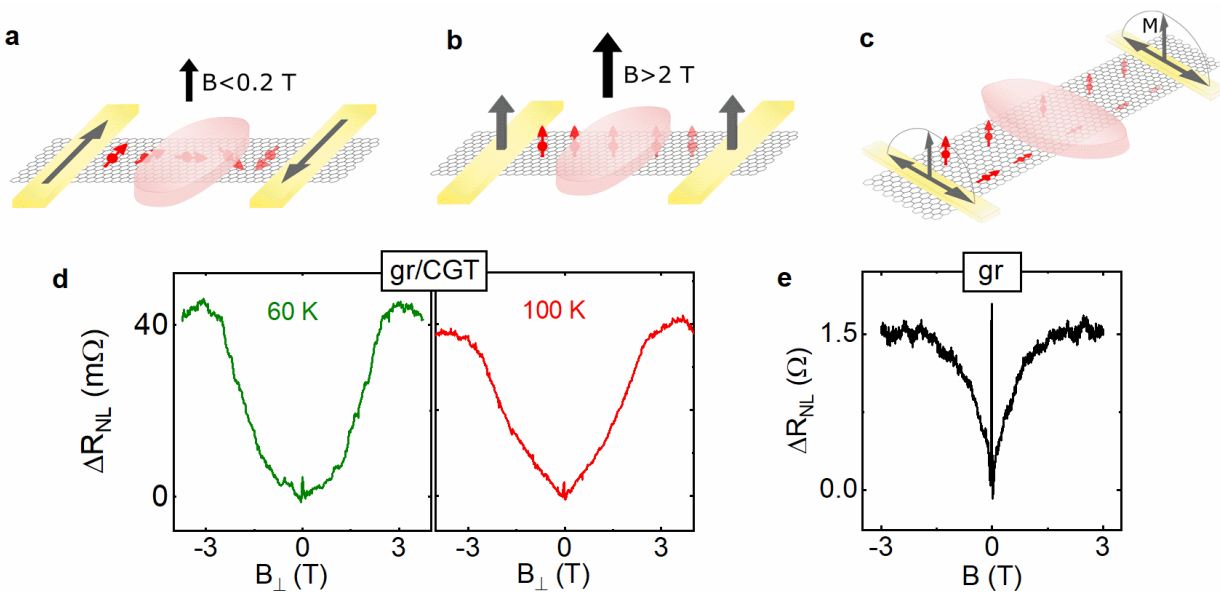

*Figure 4. Anisotropic spin relaxation in graphene-Cr$_2$Ge$_2$Te$_6$ heterostructure. a,* Measurement geometry with small perpendicular magnetic field $B_\perp<0.2$ T, where the magnetization of the ferromagnetic Co contacts and injected spin polarization are in-plane, resulting in Hanle spin precession. *b,* Measurement geometry with large perpendicular magnetic field $B_\perp>2$ T, where the magnetization of the ferromagnets and injected spin

*polarization are out of plane, resulting in no Hanle precession.* **c,** *The measurements of graphene-CGT channels indicate anisotropic relaxation with faster rates of spin relaxation for in-plane spins, as demonstrated in the schematic and* **d,** *experimental nonlocal measurements in a channel with ~60% flake overlap at 60 K and 100 K. A parabolic magnetoresistance background is subtracted from the measured data.* **e,** *The reference measurement in pristine CVD graphene showing very low anisotropy with almost similar signal for in-plane and perpendicular spins.*

**Summary and Outlook**

In summary, the observed modulation of the spin transport signal demonstrates the existance of an exchange interaction in graphene proximity-coupled to the layered magnetic insulator CGT. This is revealed by the observation of nontrivial features in the measured nonlocal Hanle spin precession signals in the graphene-CGT channel. Namely, at low temperatures there is a hysteretic shift of the Hanle signal maxima for opposite magnetic field sweep directions, accompanied by a corresponding asymmetry in the Hanle signal. These features persist up to at least 100 K, above the $T_c$ of bulk CGT. Supporting simulations reveal that these features can be explained by out-of-plane magnetism induced in the graphene layer by CGT. Additionally, measurements of large spin lifetime anisotropy indicate that proximity coupling of the graphene channel to CGT can modify its spin texture. This anisotropy can arise from the magnetic exchange induced in the graphene, while DFT simulations suggest that spin-orbit coupling could also be playing a role. Looking ahead, the proximity-induced effects can be further tuned by tailoring the van der Waals gap between graphene and CGT[30], and by placing it on both sides of the graphene layer. These achievements are important for the realization of quantum states of matter characterized by edge states which are topologically protected from backscattering without the application of an external magnetic field, namely, the quantum anomalous Hall effect[30–32]. This holds great potential for applications in spintronics and low-power quantum electronics.

**Methods**

<u>Device fabrication and measurement</u>: To fabricate the devices, CVD graphene was wet-transferred onto a Si/SiO$_2$ substrate (from Graphenea). After patterning the graphene into stripes of typical lateral width of 1-4 μm by photolithography and annealing in an Ar/H$_2$ atmosphere, 2D flakes (~30 nm thick) of the hexagonal ferromagnetic insulator Cr$_2$Ge$_2$Te$_6$ (from HQ Graphene), prepared by exfoliation method, were dry transferred immediately to reduce oxidation from air exposure. Suitable graphene-CGT heterostructure areas were chosen by optical microscope for further fabrication of ferromagnetic contacts by electron beam lithography and electron beam evaporation of metals (1 nm TiO$_2$/65 nm Co). The magnetic field sweeps at different temperatures were carried out in a Quantum Design Physical Property Measurement System while electrical biases were applied and measured by an external Keithley 6221 current source and a Keithley 2182A nanovoltmeter, respectively.

DFT simulations: For calculation of the electronic structure and structural relaxation of graphene on CGT density functional theory (DFT)[46] was used within Quantum ESPRESSO[47] using a previously developed model (ref. [31]) with a *k*-point sampling of 6x6x1 , a Hubbard parameter of *U*=1 eV[17], an energy cutoff for charge density of 500 Ry, the kinetic energy cutoff for wave functions of 60 Ry[48,49]. When SOC is included, we used the relativistic versions of the pseudopotentials. For the relaxation of the heterostructures, we added van-der-Waals corrections[50,51] and used a quasi-Newton algorithm based on a trust radius procedure. A vacuum gap of 20 Å was used to simulate quasi-2D systems. To determine the interlayer distances, the atoms of graphene were allowed to relax only in *z* direction (vertical to the layers) for determination of the interlayer distances, while the atoms of CGT were free to move in all directions until reaching a state when all components of all forces were reduced below $10^{-3}$ [Ry/$a_0$] ($a_0$ is the Bohr radius). For more details, see ref. [31].

**Acknowledgements**
Authors from Chalmers, ICN2, and University of Regensburg acknowledge funding from the European Union's Horizon 2020 research and innovation programme under grant agreement no. 785219 (Graphene Flagship Core 2). The authors at Chalmers acknowledge financial support from EU FlagEra project (from Swedish Research Council VR No. 2015-06813), Swedish Research Council VR project grants (No. 2016-03658), Graphene center and the AoA Nano program at Chalmers University of Technology. We acknowledge Dr. Ron Jansen for useful discussions about proximity induced exchange and stray Hall effects in the heterostructures. We acknowledge help from Bing Zhao in our group and staff at Nanofabrication laboratory and Quantum device laboratory for useful discussions and help in fabrication and measurements of devices. ICN2 is funded by the CERCA Programme/Generalitat de Catalunya and is supported by the Severo Ochoa program from Spanish MINECO (Grant No. SEV-2017-0706). M. Vila acknowledges funding from "La Caixa" Foundation.

**Contributions**
SPD and BK conceived the idea and designed the experiments. BK fabricated and measured the devices. AWC, MV, SR performed the simulations. KZ and JF performed the theoretical calculations. DK, AMH, AD helped in device fabrication and measurements. PS performed SQUID magnetic characterization of bulk CGT crystal.  BK, AWC, MV, KZ, JF, SR, SPD contributed in analysis of the data, compiled the figures and wrote the manuscript. All authors gave input on interpretation of the data and contributed in writing of the manuscript. SPD supervised and managed the project.

**Competing interests**
The authors declare no competing financial interests.

**Corresponding author**: saroj.dash@chalmers.se

# Supplementary information

# Magnetic proximity in a van der Waals heterostructure of magnetic insulator and graphene

Bogdan Karpiak[1], Aron W. Cummings[2], Klaus Zollner[3], Marc Vila[2,4], Dmitrii Khokhriakov[1], Anamul Md Hoque[1], André Dankert[1], Peter Svedlindh[5], Jaroslav Fabian[3], Stephan Roche[2,6], Saroj P. Dash[1,7]

[1] Department of Microtechnology and Nanoscience, Chalmers University of Technology, SE-41296, Göteborg, Sweden
[2] Catalan Institute of Nanoscience and Nanotechnology (ICN2), CSIC and The Barcelona Institute of Science and Technology, Campus UAB, Bellaterra, 08193 Barcelona, Spain
[3] Institute for Theoretical Physics, University of Regensburg, 93040 Regensburg, Germany
[4] Department of Physics, Universitat Autònoma de Barcelona, Campus UAB, Bellaterra, 08193 Barcelona, Spain
[5] Division of Solid-State Physics, Department of Engineering Science, Uppsala University, Box 534, SE-75121 Uppsala, Sweden
[6] ICREA—Institució Catalana de Recerca i Estudis Avançats, 08010 Barcelona, Spain
[7] Graphene center, Chalmers University of Technology, SE-41296, Göteborg, Sweden

Corresponding author: saroj.dash@chalmers.se

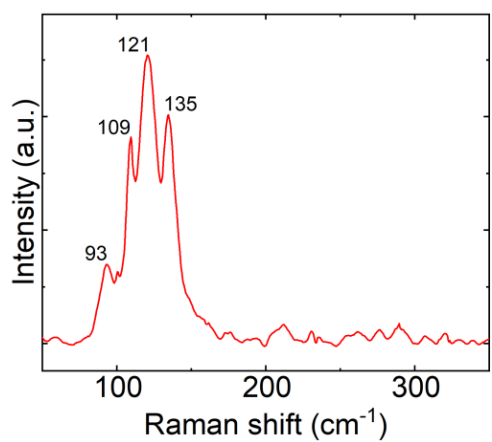

**Supplement Figure S1. Raman spectrum of the oxidized bulk CGT crystal.** The pronounced wide peak at 121 cm$^{-1}$ indicates the formation of oxides, particularly TeO$_x$[1]. The data was measured after prolonged exposure of CGT to ambient air. These Raman features are not present in the CGT flakes just after exfoliation, which are used in the device presented in the main manuscript.

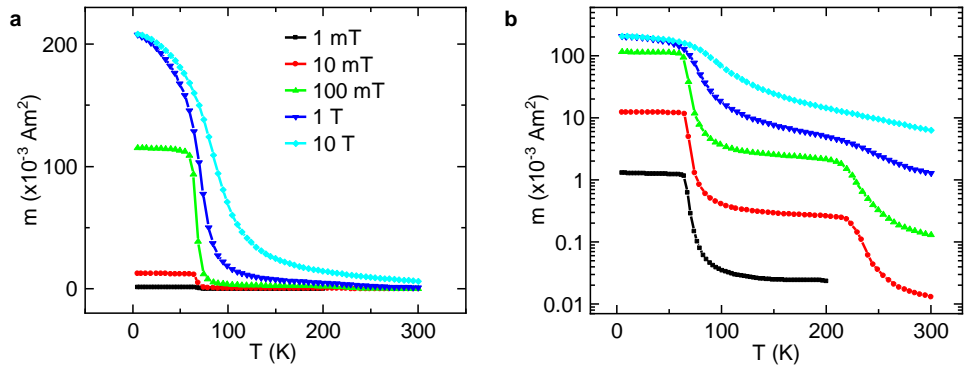

**Supplement Figure S2. Magnetic characterization of CGT. a,** Magnetic moment of bulk CGT crystal measured as a function of temperature with different applied magnetic fields showing two magnetic ordering temperatures at ~65 K and ~204 K. **b,** Same as in **a** but with log vertical scale.

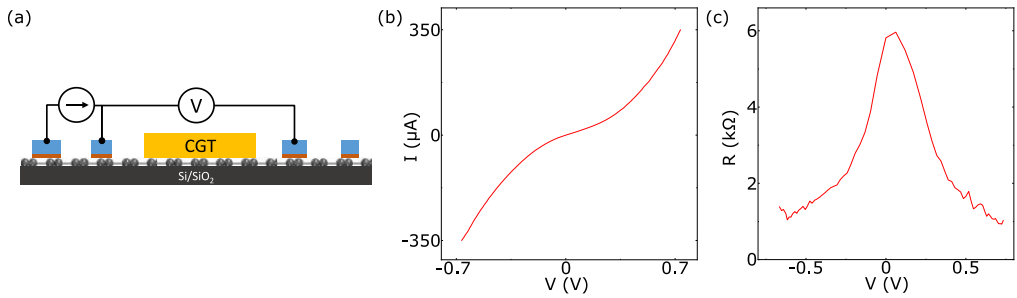

**Supplement Figure S3. Characterization of ferromagnetic tunnel contacts. a,** Schematic of 3-terminal measurements of ferromagnetic tunnel junction (TiO$_2$/Co) on graphene. **b,** Current-voltage tunneling characteristic and **c**, corresponding tunnel resistance as a function of voltage bias.

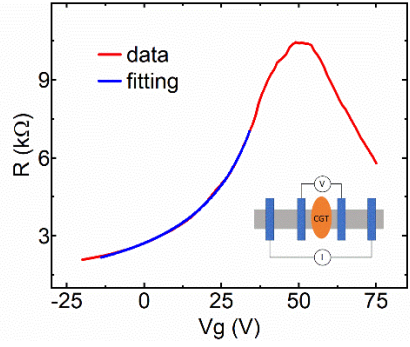

**Supplement Figure S4.** Back gate dependence of graphene channel resistance measured in a 4-terminal local geometry (measurement configuration in the inset) along with fitting in a typical graphene-CGT heterostructure device.

To extract the field-effect graphene mobility in the heterostructure channel, the gate dependence of the 4-terminal graphene local resistance (Supplement Fig. S4) was measured and fitted with

$$R = \frac{L}{W\mu \sqrt{q^2 n_0^2 + \left(\frac{(V_g - V_D)\varepsilon\varepsilon_0}{t_{ox}}\right)^2}}.$$

Here $L$, $W$ – channel width and length, $\mu$ – mobility, $q$ – electron charge, $n_0$ – residual carrier concentration, $V_g$ – gate voltage, $V_D$ – Dirac point, $\varepsilon$ – dielectric constant of SiO$_2$, $\varepsilon_0$ – permittivity of vacuum, $t_{ox}$ – SiO$_2$ thickness. From the fitting the value of graphene field effect mobility was found to be $2390 \pm 25$ cm$^2$V$^{-1}$s$^{-1}$.

# Supplementary note 1

## Simulation of the Hanle curves for graphene-CGT channel

To simulate the Hanle curves shown in Fig. 2f of the main manuscript, we use a solution of the Bloch equation that considers spatially inhomogeneous spin transport properties due to the influence of the CGT flake, which was derived in supplementary material Ref.[2]. The expression is given by

$$V_{\text{NL}} = C \cdot \text{Re} \left\{ \frac{\lambda e^{-(L_{\text{ch}} - L_{\text{H}})/\lambda}}{\frac{(\lambda_{\text{H}} + \lambda)^2}{\lambda_{\text{H}} \lambda} e^{L_{\text{H}}/\lambda_{\text{H}}} - \frac{(\lambda_{\text{H}} - \lambda)^2}{\lambda_{\text{H}} \lambda} e^{-L_{\text{H}}/\lambda_{\text{H}}}} \right\}, \tag{S1}$$

where $\lambda = \sqrt{D\tau/(1 + i\omega\tau)}$ is the spin relaxation length in the uncovered graphene region, modified by the magnetic field, with $\tau$ and $D$ the spin lifetime and spin diffusion coefficient, respectively. The Larmor precession frequency is $\omega = \gamma B_\perp$, where $\gamma$ is the gyromagnetic ratio and $B_\perp$ is the applied perpendicular magnetic field. In the graphene-CGT region, these parameters are given by $\lambda_{\text{H}} = \sqrt{D_{\text{H}} \tau_{\text{H}}/(1 + i\omega_{\text{H}} \tau_{\text{H}})}$, where $\omega_{\text{H}} = \gamma(B_\perp + B_{\text{ex}})$ and $B_{\text{ex}}$ is the perpendicular exchange field induced in the graphene by the CGT. The channel length is given by $L_{\text{ch}}$ and the length of the graphene-CGT region is $L_{\text{H}}$. In Fig. 2f and in Supplement Fig. S5a, we use the following parameters: $L_{\text{ch}} = 6.95$ μm, $L_{\text{H}} = 4.3$ μm, $\tau = 340$ ps, $\tau_{\text{H}} = 100$ ps, $D = D_{\text{H}} = 0.05$ m$^2$/s, and $C = 3.4$ mΩ. The form of $B_{\text{ex}}$ as a function of $B_\perp$ is shown in the inset of panel **a**.

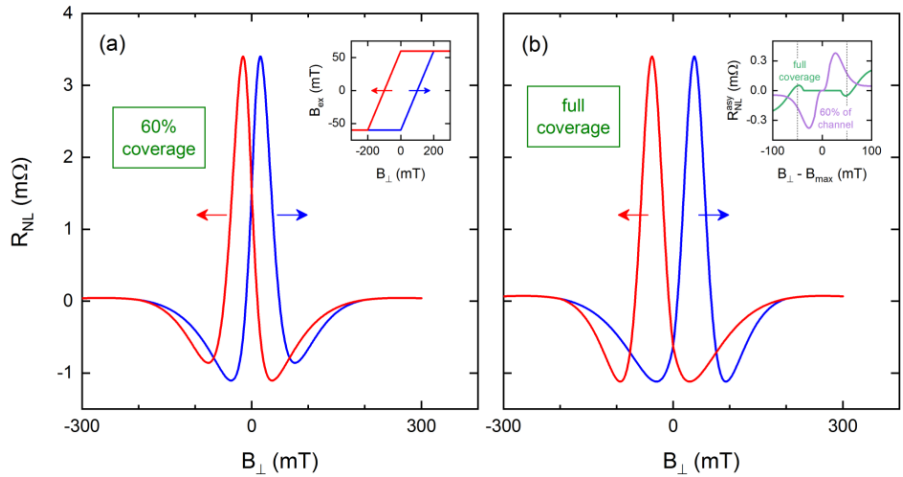

**Supplement Figure S5. Simulated Hanle curves using the parameters of Fig. 2f of the main text (and described below). a,** The case when the CGT flake covers ~60% of the graphene channel, with the inset showing the assumed behavior of B$_{\text{ex}}$. **b,** The case when the CGT flake uniformly covers the entire graphene flake. The inset shows the asymmetric portion of the forward sweep curve for each case, indicating that asymmetry in the central Hanle peak arises from the nonuniformity of the proximity-induced exchange field.

In Supplement Fig. S5, panel **a** is the same as in Fig. 2f, with the CGT covering 60% of the graphene channel. Meanwhile, panel **b** shows the case when the CGT uniformly covers the entire graphene flake. The inset shows the asymmetric component of the forward sweep in each case. Here it is evident that the asymmetry of the central Hanle peak (between the dashed lines in the inset) arises from the finite extent of the proximity-induced exchange field, and this asymmetry disappears in the limit of full coverage by the CGT.

## Supplementary note 2

### The impact of stray fields in the CGT/Gr heterostructures

The total magnetic field that shapes the Hanle signal consists of several components $B_{total} = B_{ext} + B_s + B_{exch}$, where $B_{ext}$ – externally applied field, $B_s$ – stray field, $B_{exch}$ - exchange fields. For the CGT flake with perpendicular magnetic anisotropy, one can expect $B_s$ to have direction perpendicular to the graphene channel adding to the total field. Since both the exchange and stray fields would have similar hysteresis loops reflecting on the hysteresis of the magnetization of the CGT, the impact of $B_s$ and $B_{exch}$ would be similar in the way they would shape the Hanle curve, particularly the shifts of the Hanle peaks. However, the extent of impact of these two fields could be different.

One could expect a different scaling of the relative contribution of $B_s$ and $B_{exch}$ fields on spin precession in the channel with the overlap area. To elucidate the relative impact of the stray fields on spin precession, perpendicular component of magnetic stray fields from the CGT flake $B_s$ was calculated in the middle of the channel along Y ($y = 0$) at distance of 0.35 nm from the flake in Z direction for different flake lengths along X (Fig. S6b). Considering the distribution of a perpendicular component of stray fields along the channel direction, the impact of $B_s$ is more significant only near the edges of the flake (characteristic width $\delta$), although its effect on spin precession is opposite for different sides of the same edge of the flake since $B_s$ changes sign. In the middle area of the CGT/gr channel the amplitude is small and almost constant, increasing for smaller flake lengths. However, even for flake length $L = 1$ μm in the middle $B_s = 4$ mT and dropping to $B_s = 1.1$ mT for $L = 8$ μm. A more representative value of stray fields accounting for the whole area where spins propagate is integrated average over the channel area, extended by spin diffusion length $\lambda$ at both sides of the channel

$$B_{s,eff}(L) = \int_{-L_0/2-\lambda}^{L_0/2+\lambda} \int_{-w/2}^{w/2} B_{s,L}(x,y)dxdy/(wL_0 + 2w\lambda),$$

where $B_{s,L}$ is $B_s$ calculated for flake length $L$, $L_0 = 8$ μm is the channel length and $\lambda = 2.1$ μm as obtained from Hanle signal fitting for the device described in the main text. In a similar way effective value of exchange field can be calculated as integrated average over the same area $B_{exch,eff} = B_{exch}L/(L_0 + 2\lambda)$. Here $B_{exch} = 30$ mT is taken as a constant since its value depends only on the magnetization of the flake and interface properties between graphene and CGT. The magnitude of $B_{exch}$ is taken in accordance with our estimates (Supplementary note 4). Dependence of $B_{eff}$ as a function of L (Supplement Fig. S6c) for stray fields (red) and exchange fields (blue) shows that effective contribution of exchange field is at least 10 times bigger than from stray fields even for the lower limit of the estimate of $B_{exch} = 30$ mT. Thus, one can conclude that the observed Hanle signals are mainly shaped by the applied external and induced exchange fields in the graphene channel.

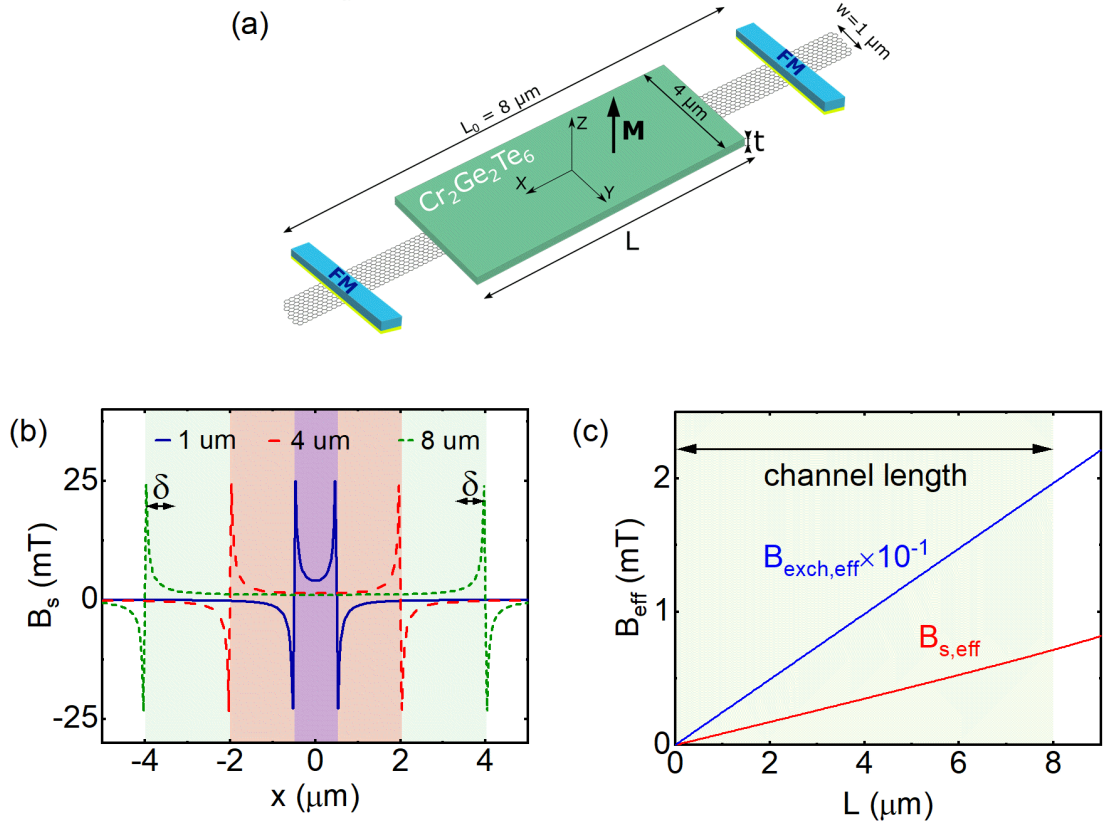

**Supplement Figure S6. The impact of stray fields in CGT/Gr heterostructures. a,** Schematic of the device with CGT/Gr heterostructure channel. **b,** Calculated perpendicular component $B_s$ of magnetic stray fields for bar magnet of width 4 µm, thickness 30 nm and variable length $L$ of 1 µm (solid blue), 4 µm (dash red) and 8 µm (short dash green) with magnetization M=164 kA/m at distance of 0.35 nm from the magnet along the Z axis[3,4]. The amplitude of $B_s$ quickly drops with characteristic distance from the edge $\delta$ (marked for 8 µm flake length). Background color marks CGT flake length along X axis, respectively. **c,** Effective value of the perpendicular component of the stray magnetic field at distance of 0.35 nm from the CGT flake averaged over the channel area, extended by spin diffusion length $\lambda$ on both sides $B_{s,eff}(L) = \int_{-L_0/2-\lambda}^{L_0/2+\lambda} \int_{-w/2}^{w/2} B_{s,L}(x,y)dxdy/(wL_0 + 2w\lambda)$ for device dimensions as in **a** (red) and effective value of exchange field averaged over the same area $B_{exch,eff} = B_{exch}L/(L_0 + 2\lambda)$ for $B_{exch} = 30$ mT and divided by 10 (blue) as a function of flake length L along the X axis.

# Supplementary note 3

**Asymmetry and Hanle peaks separation change with varying overlap of the channel by the CGT flake**

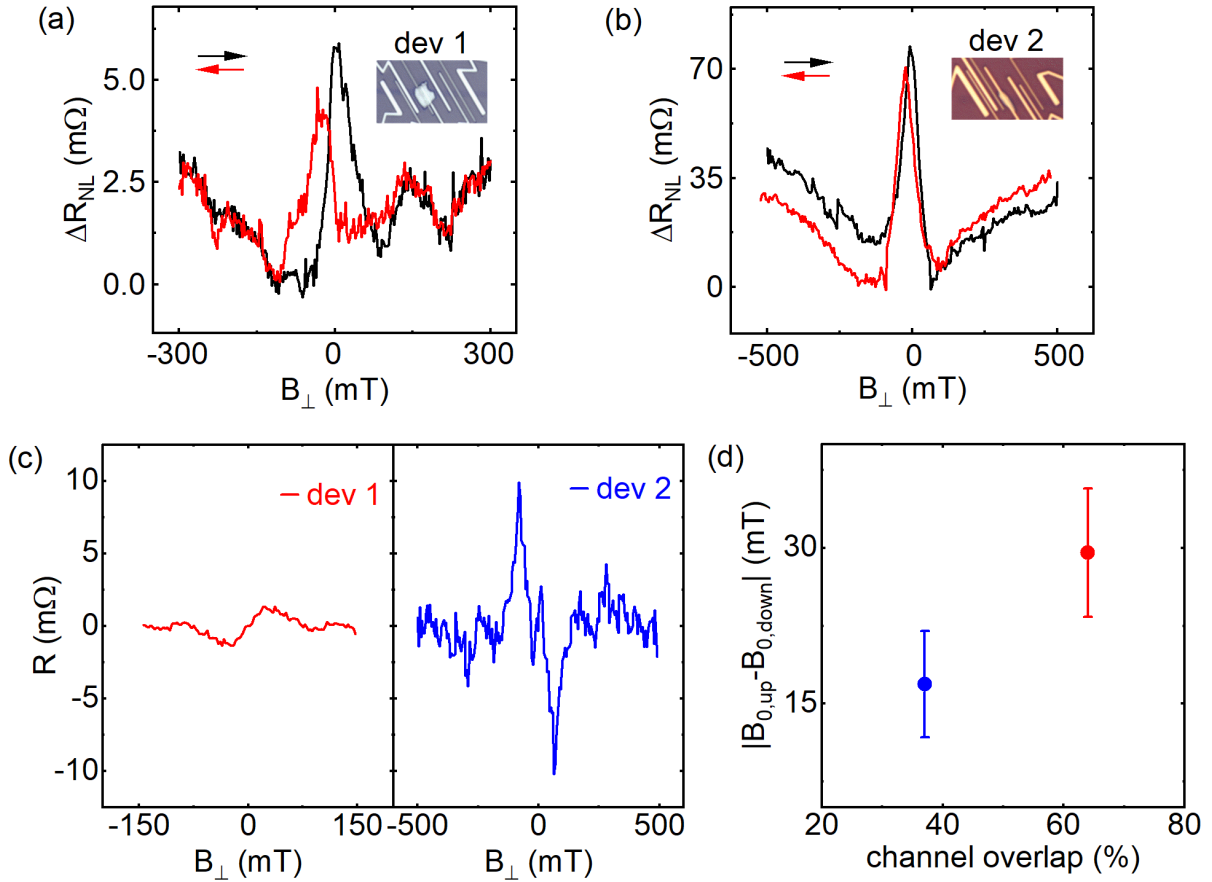

**Supplement Figure S7. The change in asymmetry and Hanle peak separation between different CGT/Gr heterostructure channels with varying overlap by the CGT flake. a,b** Nonlocal Hanle spin precession signals at 50 K for two devices with CGT-Gr overlap areas of 64% (dev 1) and 37% (dev2) and channel lengths of 6.9 $\mu m$ and 4 $\mu m$, respectively. Insets: optical microscope pictures of these devices, respectively. **c,** Asymmetric part of the measured Hanle signals at 50 K in dev1 and dev2. **d,** Magnitude of the horizontal separation between positions of nonlocal Hanle peaks for up ($B_{0,up}$) and down ($B_{0,down}$) sweeps as a function of channel overlap by the CGT flake for dev 1 (red) and dev 2 (blue), respectively.

The nonlocal Hanle spin precession signals were measured in two devices with CGT/Gr heterostructures channels with channel overlaps by the CGT flake of 64 % (dev 1) and 37 % (dev 2) as shown in Supplement Figure S7a,b. From the asymmetrical part of the Hanle signals in the two devices, obtained after deconvolution of the raw spin precession signals[5] (Supplement Fig. S7c), a decrease in asymmetry is observed with increased channel overlap. This is in agreement with our simulation results (Supplementary note 1). Additionally, a decrease in separation between Hanle peaks between forward and backward magnetic field sweeps is also observed with decrease in channel overlap (Supplement Fig. S7d) as one can expect due to smaller part of the channel influenced by the exchange fields that cause separation of Hanle peaks.

## Supplementary note 4

**Fits to the experimental Hanle curves**

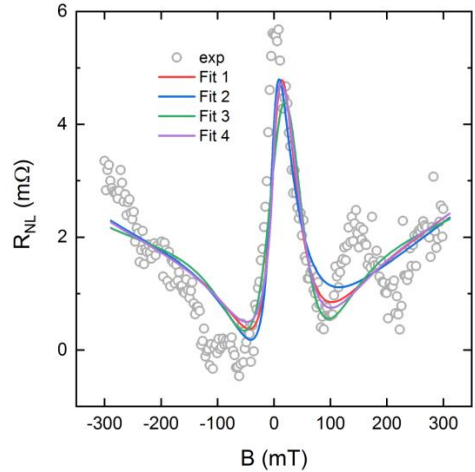

**Supplement Figure S8.** Fits to the experimental Hanle curve shown in Fig. 2e, using Eq. (S1). Here we only show the forward sweep, as the reverse sweep fits would be a mirror image.

In Supplement Fig. S8 we show four different fits to the experimental data of Fig. 2e. The fits were done using Eq. (S1) and simultaneously fitting the sum and difference of the forward and reverse magnetic field sweeps. We considered four fits under different restrictions. In Fit 1 we let all parameters vary independently, but with upper bounds $\tau, \tau_H$ < 1 ns; $D, D_H$ < 0.2 m$^2$/s; and $B_{ex}$ < 1 T. In Fit 2 we kept the same restrictions but we fixed $\tau$ = 500 ps and $D$ = 0.02 m$^2$/s, similar to previously measured CVD graphene devices[6]. In Fit 3, we forced uniform spin transport throughout the graphene channel by setting $\tau_H = \tau$ and $D_H = D$. In Fit 4, we allowed all parameters to vary independently without any upper bounds. The results are shown in the table below. Except for fit 2, all fits yield a proximity-induced exchange field on the order of a few 10's of mT.

|  | Fit 1 | Fit 2 | Fit 3 | Fit 4 |
|---|---|---|---|---|
| $\tau$ (ps) | 351 | 500 | 121 | 1990 |
| $\tau_H$ (ps) | 105 | 10 | " | 106 |
| $D$ (m$^2$/s) | 0.059 | 0.02 | 0.042 | 0.246 |
| $D_H$ (m$^2$/s) | 0.073 | 0.11 | " | 0.04 |
| $B_{ex}$ (mT) | 58 | 209 | 42 | 33 |

# Supplementary note 5

## DFT simulations of graphene-CGT heterostructure

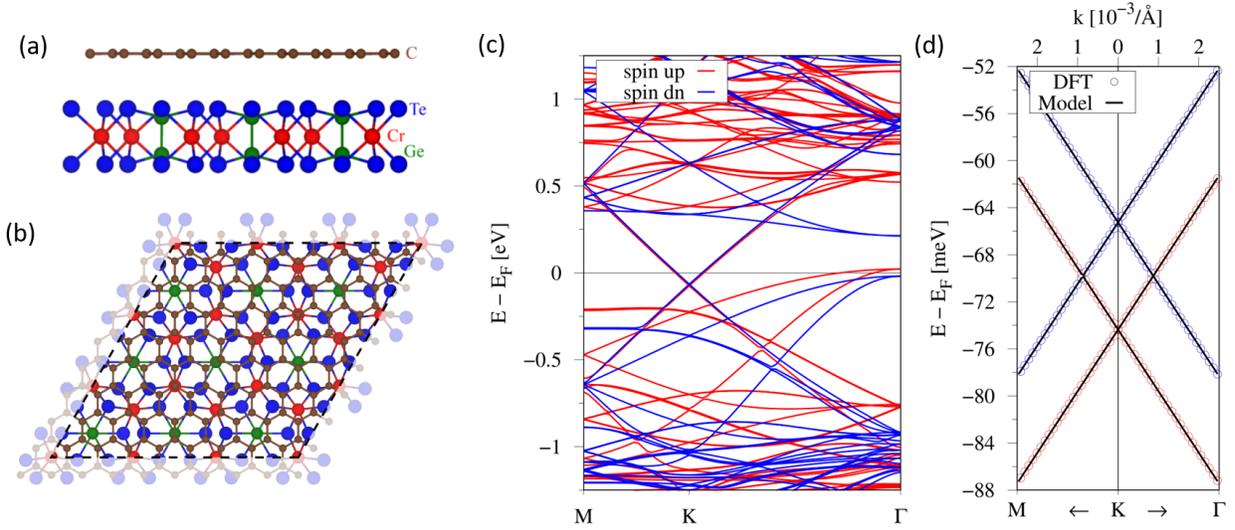

**Supplement Figure S9. DFT simulations of graphene-CGT heterostructure for a system of 218 atoms without SOC. a,** Side view and **b,** top view of the simulated atomic structure of graphene on monolayer CGT (supercell). **c,** Calculated electronic band structure of graphene without SOC. Zoom to the fine structure of the low energy bands near the charge neutrality point. Bands with spin up (down) are shown in red (blue). Symbols are first principles data and solid lines are the fits using the low-energy graphene model described below.

The simulated heterostructure of graphene on monolayer CGT is shown in Supplement Fig. S9a,b, where an 8x8 supercell of graphene is placed on a $3 \times 3$ CGT supercell. We stretch the lattice constant of graphene from $a$ = 2.46 Å to 2.50 Å and compress the lattice constant of CGT from 6.8275 Å[3] to 6.666 Å. The resulting supercell has a lattice constant of 20.0 Å and contains 218 atoms in the unit cell.

Supplement Fig. S9c shows the resulting band structure of graphene without SOC and Supplement Fig. S9d zoom to the fine structure near the charge neutrality point. The symbols show the DFT results, and their color indicates the direction of out-of-plane spin polarization of the bands (up or down). The solid lines are the fits using a low-energy model of the graphene bands in the vicinity of the Dirac points,

$$\mathcal{H} = \mathcal{H}_0 + \mathcal{H}_\Delta + \mathcal{H}_{\text{ex}} + E_{\text{D}},$$

$$\mathcal{H}_0 = \hbar v_{\text{F}}(\tau k_x \sigma_x - k_y \sigma_y) \otimes s_0,$$

$$\mathcal{H}_\Delta = \Delta \sigma_z \otimes s_0,$$

$$\mathcal{H}_{\text{ex}} = \left(-\lambda_{\text{ex}}^{\text{A}} \sigma_+ + \lambda_{\text{ex}}^{\text{B}} \sigma_-\right) \otimes s_z,$$

with the Fermi velocity $v_F$; the Cartesian components of the wave vector $k_x$ and $k_y$ measured from K/K'; the valley index $\tau = \pm 1$ for K/K'; the Pauli spin matrices $s_i$ acting on spin space (↑,↓) and the pseudospin matrices $\sigma_i$ acting on sublattice space (A,B), with $i = \{0, x, y, z\}$ and $\sigma_\pm = (\sigma_z \pm \sigma_0)/2$; and the staggered potential gap $\Delta$. The second-to-last term describes the sublattice-resolved proximity exchange with parameters $\lambda_{ex}^A$ and $\lambda_{ex}^B$. The four basis states we use are $|\Psi_A, \uparrow\rangle$, $|\Psi_A, \downarrow\rangle$, $|\Psi_B, \uparrow\rangle$, $|\Psi_B, \downarrow\rangle$. Note that the model Hamiltonian is centered around zero energy Fermi level. For the first principles results this is not necessarily the case. Therefore, we introduce another parameter $E_D$, which we call the Dirac point energy, and which introduces a shift of the global band structure.

Fits of this model to the DFT simulations yield the following set of parameters: $v_F = 7.950 \times 10^5$ m/s, $\Delta = 0.006$ meV, $\lambda_{ex}^A = 4.556$ meV, $\lambda_{ex}^B = 4.558$ meV, and $E_D = -70$ meV. Combining the sublattice-dependent terms, this yields a ferromagnetic exchange splitting of $\lambda_{ex} = (\lambda_{ex}^A + \lambda_{ex}^B)/2 = 4.557$ meV.

To obtain an estimate of SOC strength in the graphene proximitized by CGT, we performed similar DFT simulations, but with inclusion of SOC and in smaller system (Fig. S10). The simulated heterostructure of graphene on monolayer CGT is shown in Supplement Fig. S10a,b, where a 5x5 supercell of graphene is placed on a $\sqrt{3} \times \sqrt{3}$ CGT supercell. We keep the lattice constant of graphene unchanged at $a$ = 2.46 Å and stretch the lattice constant of CGT by roughly 4% from 6.8275 Å[3] to 7.1014 Å. The resulting supercell has a lattice constant of 12.3 Å and contains 80 atoms in the unit cell. Theoretical calculations predict that the tensile strain leaves the ferromagnetic ground state unchanged, but enhances the band gap and the Curie temperature of CGT[7,8]. The average distance between graphene and CGT was relaxed to 3.516 Å, consistent with literature[9].

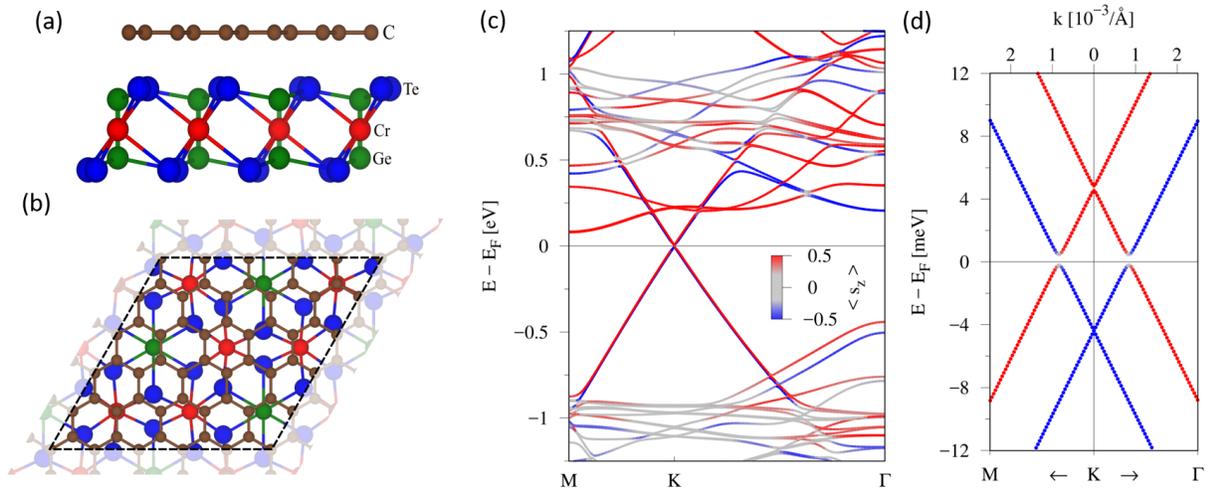

**Supplement Figure S10. DFT simulations of graphene-CGT heterostructure for a system of 80 atoms with SOC. a,** Side view and **b,** top view of the simulated atomic structure of graphene on monolayer CGT. **c,** Band structure of the graphene layer and **d,** zoom to the fines structure near the charge neutrality point.

Symbols are DFT results, whose color indicates the out-of-plane spin polarization of the bands. Solid lines (overlap with the symbols) are the fits using the low-energy graphene model described below.

Supplement Fig. S10c shows the resulting band structure of graphene along with zoomed region near the charge neutrality point (Supplement Fig. S10d). The symbols show the DFT results, and their color indicates the degree of out-of-plane spin polarization of the bands. The solid lines are the fits using a low-energy model of the graphene bands in the vicinity of the Dirac points at K and K′,

$$\mathcal{H} = \mathcal{H}_0 + \mathcal{H}_\Delta + \mathcal{H}_\mathrm{I} + \mathcal{H}_\mathrm{R} + \mathcal{H}_\mathrm{ex} + E_\mathrm{D},$$

$$\mathcal{H}_\mathrm{I} = \tau(\lambda_\mathrm{I}^\mathrm{A}\sigma_+ + \lambda_\mathrm{I}^\mathrm{B}\sigma_-)\otimes s_z,$$

$$\mathcal{H}_\mathrm{R} = \lambda_\mathrm{R}(\tau\sigma_x\otimes s_y + \sigma_y\otimes s_x),$$

with the spin-orbit parameters $\lambda_\mathrm{I}^\mathrm{A}$ and $\lambda_\mathrm{I}^\mathrm{B}$ for the sublattice-resolved intrinsic SOC and $\lambda_\mathrm{R}$ for the Rashba SOC.

Fits of this model (with SOC) to the DFT simulations yield the following set of parameters: $v_\mathrm{F} = 8.176 \times 10^5$ m/s, $\Delta = 0.105$ meV, $\lambda_\mathrm{R} = 0.253$ meV, $\lambda_\mathrm{I}^\mathrm{A} = 0.123$ meV, $\lambda_\mathrm{I}^\mathrm{B} = -0.102$ meV, $\lambda_\mathrm{ex}^\mathrm{A} = -4.490$ meV, $\lambda_\mathrm{ex}^\mathrm{B} = -4.321$ meV, and $E_\mathrm{D} = 0.015$ meV. Combining the sublattice-dependent terms, this yields a ferromagnetic exchange splitting of $\lambda_\mathrm{ex} = (\lambda_\mathrm{ex}^\mathrm{A} + \lambda_\mathrm{ex}^\mathrm{B})/2 = -4.406$ meV and a valley-Zeeman SOC of $\lambda_\mathrm{VZ} = (\lambda_\mathrm{I}^\mathrm{A} - \lambda_\mathrm{I}^\mathrm{B})/2 = 0.113$ meV. The values of $\lambda_\mathrm{R}$ and $\lambda_\mathrm{VZ}$ can then be used to estimate the spin lifetime anisotropy under the assumption that all spin relaxation in graphene arises from SOC; see the section "Anisotropic spin relaxation in graphene-CGT insulator heterostructures" in the main text.

One can notice, when comparing the big supercell (218 atoms) with the small supercell (80 atoms) calculations, that the proximity exchange switches sign. The magnitude of the proximity exchange is roughly unchanged. This is quite surprising and could arise due to twist angle, which changes with the supercell size. It has been shown before that the twist angle can change SOC in graphene-TMDC heterostructures[10]. Considering this, graphene-CGT heterostructures is not trivial in terms calculation of proximity exchange and SOC and requires a more detailed separate study.

## Supplementary note 6

### Spin relaxation due to exchange fluctuations

When graphene possesses an effective magnetic exchange field $B_{\text{ex}}$, either locally by adatom adsorption[11] or by proximity to a magnetic substrate[12], random fluctuations of $B_{\text{ex}}$ can relax the spins in the graphene layer. Importantly, if an external magnetic field $B_{\text{ap}}$ is applied, the magnetic moments of the magnetic material will align with it and so will the exchange fields induced in graphene. This will cause a suppression of the relaxation because, although the strength of the fluctuations is given primarily by the temperature, their relative strength compared to the exchange field component parallel to $B_{\text{ap}}$ will diminish. This effect is different for spins parallel and perpendicular to $B_{\text{ap}}$, and it is this difference that generates a spin lifetime anisotropy. In the absence of both $B_{\text{ap}}$ and $B_{\text{ex}}$ (i.e., a random distribution of paramagnetic moments), the anisotropy is $r \equiv \tau_\perp/\tau_\parallel = 1$. When the moments start to align due to presence of either $B_{\text{ap}}$ or $B_{\text{ex}}$, the spins parallel to that component will relax more slowly and generate anisotropy. If that component is perpendicular to the graphene plane, then anisotropy $r > 1$ is expected. The following equations for the spin relaxation rate of in-plane $(1/\tau_\parallel^{\text{ex}})$ and out-of-plane $(1/\tau_\perp^{\text{ex}})$ spins, assuming $B_{\text{ap}}$ and $B_{\text{ex}}$ are perpendicular to the graphene plane, summarize this mechanism (hereafter we let $B_{\text{ap}} = B_\perp$ for consistency with the main text):

$$\frac{1}{\tau_\parallel^{\text{ex}}} = \frac{\Gamma}{2} + \frac{\Gamma}{2}\frac{\gamma^2}{(B_\perp + B_{\text{ex}})^2 + \gamma^2},$$

$$\frac{1}{\tau_\perp^{\text{ex}}} = \Gamma\frac{\gamma^2}{(B_\perp + B_{\text{ex}})^2 + \gamma^2},$$

with $\gamma = \frac{\hbar}{g\mu_B \tau_c}$ and $\Gamma = \frac{\Delta B^2}{\tau_c \gamma^2}$, where $g$ is the electron g-factor, $\mu_B$ is the Bohr magneton, $\tau_c$ is the correlation time of the exchange fluctuations (i.e., the time for an electron to experience a different exchange field) and $\Delta B$ is the strength of the exchange fluctuations.

To show how the measurement of Fig. 4d can be explained by exchange fluctuations, we first assume that this mechanism is solely responsible for spin relaxation in graphene under the CGT. It is important to remember that in Fig. 4d, the anisotropy $r = 3.9$ was extracted from the ratio of the nonlocal spin signals at $B_\perp = 0$ and $B_\perp = 2.5$ T. Plugging this into the above equations gives

$$r = \frac{\tau_\perp^{\text{ex}}(B_\perp = 2.5\text{ T})}{\tau_\parallel^{\text{ex}}(B_\perp = 0)} = \frac{\frac{\Gamma}{2} + \frac{\Gamma}{2}\frac{\gamma^2}{B_{\text{ex}}^2 + \gamma^2}}{\Gamma\frac{\gamma^2}{(2.5 + B_{\text{ex}})^2 + \gamma^2}} \approx \frac{\frac{\Gamma}{2} + \frac{\Gamma}{2}\frac{\gamma^2}{B_{\text{ex}}^2 + \gamma^2}}{\Gamma\frac{\gamma^2}{2.5^2 + \gamma^2}} \approx \frac{\frac{\Gamma}{2} + \frac{\Gamma\gamma^2}{2\gamma^2}}{\Gamma\frac{\gamma^2}{2.5^2 + \gamma^2}} = \left(\frac{2.5}{\gamma}\right)^2 + 1,$$

where we have assumed that $B_{\text{ex}} \ll 2.5$ T and $B_{\text{ex}} \ll \gamma$, since the Hanle measurements indicate values of $B_{\text{ex}}$ on the order of a few 10's of mT. Using the experimental value $r = 3.9$ gives $\gamma = 1.5$ T, yielding $\tau_c = 4$ ps (in the range of estimates used for graphene-YIG heterostructures[12]) and justifying our assumption that $B_{\text{ex}} \ll \gamma$. Finally, plugging this value of $\gamma$ into the expression for spin lifetime anisotropy at $B_\perp = 0$ yields

$$r(B_\perp = 0) = \frac{\tau_\perp^{\text{ex}}(B_\perp = 0)}{\tau_\parallel^{\text{ex}}(B_\perp = 0)} = \frac{\frac{\Gamma}{2} + \frac{\Gamma}{2}\frac{\gamma^2}{B_{\text{ex}}^2 + \gamma^2}}{\Gamma \frac{\gamma^2}{B_{\text{ex}}^2 + \gamma^2}} = 1 + \frac{1}{2}\left(\frac{B_{\text{ex}}}{\gamma}\right)^2 \approx 1.002,$$

where we have assumed a generous value of $B_{\text{ex}} = 100$ mT. This result indicates that if the spin relaxation is dominated by exchange fluctuations, the spin lifetime anisotropy at zero external magnetic field is isotropic, owing to the relatively small value of $B_{\text{ex}}$, and that the large anisotropy seen in Fig. 4d of the main text is driven by the external field $B_\perp$ and is not intrinsic to the graphene/CGT interface.

**Supplementary material references**